# On the Theory of Vibronic Laser

**A. P. Saiko**

*Institute of Solid State and Semiconductor Physics, National Academy of Belarus, Minsk, 220072 Belarus*
*e-mail: saiko@ifttp.bas-net.by*

**Abstract**—Some aspects of lasing at vibronic transitions in impurity crystals are theoretically studied. The threshold conditions for a vibronic laser are shown to be dependent on the strength of interaction of optical centers with a local vibration, which forms the vibronic spectrum, and the crystal lattice temperature. The theory can be easily generalized to the spectrum containing a structureless phonon sideband and well agrees with the experimental temperature dependence of the output power of a $Mg_2SiO_4$: $Cr^{4+}$ forsterite laser.

## 1. INTRODUCTION

The development of tunable solid-state lasers is one of the most urgent scientific and practical problems of laser physics, which has been actively elaborated in the last years.

The lasing conditions in solids are significantly complicated because of a strong adiabatic vibronic interaction, which determines the structure of vibronic levels, the main lasing parameters being dependent on the temperature of the laser medium and the strength of vibronic interaction (see, for example, [1-4]). In this case, a phenomenological description of lasing based on the use of the rate equations for level populations and polarization of the resonance medium, when the effect of the lattice and intramolecular vibrations is reduced only to the broadening of energy levels, cannot be considered adequate. A more fundamental, microscopic description of the laser action in vibronic systems, which explicitly takes the electron–vibrational interaction into account, will not only elucidate the fundamental problems of lasing at vibronic transitions in impurity crystals but also will be helpful in the search for new, more efficient laser materials.

The vibronic structure of optical spectra has been explicitly considered in the theoretical studies [5, 6] of threshold conditions and the radiation field dynamics of a vibronic laser performed within the framework of semiclassical equations of motion for field amplitudes and generalized operators of the transition between vibronic levels.

Below, the following aspects of the theory of a vibronic laser are discussed: The microscopic approach to the derivation of laser equations is realized, the linear Fokker–Planck equation is obtained for the distribution function of the laser field amplitudes, the threshold lasing conditions are discussed [7, 8], and the theory developed in this paper is applied to the description of lasing in a chromium-doped forsterite single crystal.

## 2. THE LINEAR FOKKER–PLANCK EQUATION FOR THE DISTRIBUTION FUNCTION OF LASER FIELD AMPLITUDES

We assume that impurities in a crystal, which exhibit lasing, can be treated as two-level quantum objects, whose ground and excited states are adiabatically coupled with phonon modes of the crystal lattice and local (for example, intramolecular) vibrations, whose frequencies exceed the maximum frequency of lattice phonons. The density matrix $\rho$ of an ensemble of impurities interacting with vibrational modes and the electromagnetic field in a cavity is described by the master equation

$$\frac{\partial \rho(t)}{\partial t} = -i(L_f + L_a + L_{af} + L_l + L_{al} + L_L \\ + L_{aL} + i\Lambda_f + i\Lambda_a + i\Lambda_l)\rho(t), \quad (1)$$

where

$$L_m X \equiv [H_m, X],$$

$H_m$ and $L_m$ are Hamiltonians and corresponding Liouvillians, respectively, $m = a, f, l$, etc.,

$$H_f = \omega a^+ a, \quad H_a = \varepsilon \sum_j R_j^z$$

are Hamiltonians of the single-mode electromagnetic field (radiation is collinear to an elongated pencil-like sample) and optical two-level centers;

$$H_{af} = g \sum_j (a R_j^+ + \text{H.c.})$$

is the operator of interaction of the optical centers with the radiation field; $H_l = \nu b^+ b$ is the Hamiltonian of the intramolecular vibration;

$$H_{al} = \sum_j \xi \nu R_j^z (b + b^+)$$

is the electron–vibrational interaction that produces the vibronic structure in the optical spectrum;

$$H_L = \sum_k \omega_k c_k^+ c_k$$

is the lattice Hamiltonian;

$$H_{aL} = \sum_k \lambda_k R_j^z (c_k + c_k^+)$$

is the electron–phonon interaction operator; $a^+$, $b^+$, $c_k^+$, and $\omega$, $\nu$, $\omega_k$ are the creation operators for photons, intramolecular vibrational modes, $k$th phonon mode and the corresponding frequencies, respectively; $g$, $\xi$, and $\lambda_k$ are the interaction coefficients; $R_j^{\pm, z}$ are operators of the energy spin, which describe the $j$th two-level optical center ($j = 1, 2, \ldots, N$) and are identical to the Pauli spin matrices; Planck's constant is assumed equal to unity. Liouvillians $\Lambda_f$ and $\Lambda_a$ take into account incoherent interactions that result in the energy dissipation from the radiation field and excited optical impurities, respectively:

$$\Lambda_f X = \kappa([a(X, a^+)] + \text{H.c.}) + 2\kappa \bar{N}[a, [X, a^+]], \quad (2)$$

$$\Lambda_a X = \frac{1}{2} \sum_j \{\gamma_{12}([R_j^- X, R_j^+] + \text{H.c.}) + \gamma_{21}([R_j^+ X, R_j^-] + \text{H.c.})\} \quad (3)$$

$$+ \Gamma(T) \sum_j ([R_j^z \rho, R_j^z] + \text{H.c.}),$$

where $\kappa$ is the decay constant of the radiation field caused by the irreversible escape of photons outside an elongated sample of length $l$ (we assume that $\kappa = c/l$, where $c$ is the speed of light); $\bar{N} = [\exp(\omega/k_B T) - 1]^{-1}$; $T$ is the sample temperature; $\gamma_{12}(\gamma_{21})$ is the rate of transition from the ground (excited) to the excited (ground) electronic state of the impurity. The latter, purely dephasing term in (3) describes the electron–phonon interaction, which is bilinear in the phonon variables. For example, if dephasing is predominantly determined by a narrow part of the phonon spectrum, then, in the second-order of the perturbation theory in the electron–phonon coupling constant $\chi$, we have [9]

$$\Gamma(T) = \chi^2 \bar{n}(\omega_0)[\bar{n}(\omega_0) + 1], \quad (4)$$

where $\omega_0$ is an effective Einstein frequency of the crystal lattice and $\bar{n}(\omega_0) = [\exp(\omega_0/k_B T) - 1]^{-1}$. The Liouvillian $\Lambda_l$ describes the decay of the intramolecular vibrational mode

$$\Lambda_l X = \alpha([bX, b^+] + \text{H.c.}) + 2\alpha \bar{n}[b, [X, b^+]], \quad (5)$$

where $\alpha$ is the decay constant and $\bar{n} = [\exp(\nu/k_B T) - 1]^{-1}$.

The form and derivation of Liouvillians (2), (3), and (5) that take the dissipative processes into account are presented, for example, in papers [9, 10].

The model of an ensemble of resonance optical centers interacting with the radiation field in a crystal considered here is quite general. In particular, a practically important and often encountered case of the impurity systems whose optical spectrum consists only of one zero-phonon line and a structureless phonon sideband, which can produce lasing, is a simplified version of this model.

To derive the master equation describing the laser action, one should adiabatically exclude the variables related to the polarization ($\propto R_j^\pm$), inverse population ($\propto R_j^z$), amplitudes of local vibrations ($b$, $b^+$), and lattice modes ($c_k$, $c_k^+$), because these quantities rapidly vary in time, whereas the amplitudes $a$ and $a^+$ of the light field slowly vary in time. Using the methods of nonequilibrium statistical mechanics [11, 12] and taking into account that the orders of magnitude of the operators for a high-Q cavity satisfy the inequalities $O(\Lambda_f) \ll O(L_{af}) \ll O(\Lambda_a)$, we obtain, in the second-order approximation in $L_{af}$ (or $H_{af}$), the master equation for the density matrix $\sigma = \text{Sp}_a \text{Sp}_l \text{Sp}_L(\rho)$ of the field in the cavity

$$\dot{\sigma}(t) = \Lambda_f \sigma(t) - \int_0^1 d\tau \text{Sp}_a \text{Sp}_l \text{Sp}_L \quad (6)$$

$$\times \{\tilde{L}_{af}(t) \exp[\Lambda_a(t-\tau)] \tilde{L}_{af}(\tau) \rho_a \rho_l \rho_L \sigma(\tau)\},$$

where

$$\tilde{L}_{af}(t) = \exp(i\tilde{L}t) L_{af} \exp(-i\tilde{L}t),$$

$$\tilde{L} = L_a + L_f + L_l + L_{al} + L_L + L_{aL} + i\Lambda_l,$$

$$\rho_{l, L} = \frac{\exp(-H_{l, L}/k_B T)}{\text{Sp}_{l, L}[\exp(-H_{l, L}/k_B T)]},$$

$$\rho_a = \Pi_j \left(\frac{1}{2} + \sigma_0 R_j^z\right),$$

$\sigma_0 = (\gamma_{12} - \gamma_{21})/(\gamma_{12} + \gamma_{21})$ is the inverse population of the levels ($-1 \leq \sigma_0 \leq 1$), $\gamma_{12}(\gamma_{21})$ is the rate of transition from the ground (excited) electronic state to the excited (ground) state.

After some transformations, Eq. (6) can be rewritten in the form

$$\dot{\rho}(t) = \Lambda_f \sigma(t) - g^2 \int_0^1 d\tau \exp[-\gamma_\perp(t-\tau)] \\
\times \{\exp[i\Delta(t-\tau)]\langle U^+(t)U(\tau)\rangle_{l,L} \quad (7) \\
\times [\mathrm{Sp}_a(R^+R^-\rho_a)(aa^+\sigma(\tau) - a^+\sigma(\tau)a) \\
+ \mathrm{Sp}_a(R^-R^+\rho_a)(\sigma(\tau)a^+a - a\sigma(\tau)a)] + \mathrm{H.c.}\},$$

where $\Delta = \varepsilon - \omega$, $\gamma_\perp = (\gamma_{12} + \gamma_{21})/2$,

$$R^{\pm,z} = \sum_j R_j^{\pm,z}, \quad \langle\ldots\rangle_{l,L} = \mathrm{Sp}_l\mathrm{Sp}_L\{\ldots\rho_l\rho_L\},$$

$$\langle U^+(t)U(\tau)\rangle_{l,L} = \left\langle \overset{(+)}{T}\left(\exp\left[i\int_0^1 dt_1 F(t_1)\right]\right)\right\rangle \quad (8) \\
\times \left\langle \overset{(-)}{T}\left(\exp\left[-i\int_0^1 dt_2 F(t_2)\right]\right)\right\rangle_{l,L},$$

$$F(t) = V(t)\left[\xi\nu(b+b^+) + \sum_k \lambda_k(c_k + c_k^+)\right]V^+(t), \quad (9)$$

$$V(t) = \exp[i(H_l + H_L + H_{al} + H_{aL} + i\Lambda_l)t]. \quad (10)$$

The correlation function (8) can be calculated as in [9, 10]. Then, we have

$$\ln\langle U^+(t)U(t')\rangle_{l,L} \approx \xi^2\{-2(2\bar{n}+1) \\
- [(2\bar{n}+1)\alpha + i\nu](t-t') \\
+ [\bar{n}\exp[i\nu(t-t')] + (\bar{n}+1)\exp[-i\nu(t-t')]] \\
\times \exp(-\alpha|t-t'|)\} + \sum_k \left(\frac{\lambda_k}{\omega_k}\right)^2 \quad (11) \\
\times [-2(\bar{n}_k+1) - i\omega_k(t-t') + \bar{n}_k\exp[i\omega_k(t-t')] \\
+ (\bar{n}_k+1)\exp[-i\omega_k(t-t')]],$$

where $\bar{n}_k = [\exp(\omega_k/k_BT) - 1]^{-1}$. The coefficients $i\xi^2\nu$ and $i\sum_k \lambda_k^2/\omega_k$ in the terms linear in $(t-t')$ in the right-hand side of expression (11) will be omitted below, assuming that they have been already included into the renormalized values of energies of the ground and excited electronic states.

By passing in (7) to the P-representation of Glauber–Sudarshan for $\sigma$,

$$\sigma(t) = \int d^2\alpha P(\alpha, \alpha^*, t)|\alpha\rangle\langle\alpha|,$$

where $a|\alpha\rangle = \alpha|\alpha\rangle$, we obtain, in the Markov approximation, instead of (7) the Fokker–Planck equation

$$\dot{P}(\alpha, \alpha^*, t) = \left\{\kappa\left(\frac{\partial}{\partial\alpha}\alpha + \frac{\partial}{\partial\alpha^*}\alpha^*\right) - \frac{g^2 N\sigma_0}{\gamma_\perp} \\
\times \left(\int_0^\infty d\tau\gamma_\perp \exp[(i\Delta-\gamma_\perp)\tau]\right. \\
\times \langle U^+(t)U(\tau)\rangle_{l,L}\frac{\partial}{\partial\alpha^*}\alpha^* + \mathrm{c.c.}\right) + \\
+ \left[2\kappa\bar{N} + \frac{g^2 N(1+\sigma_0)}{2\gamma_\perp}\right. \quad (12) \\
\times \left(\int_0^\infty d\tau\gamma_\perp \exp[(i\Delta-\gamma_\perp)\tau]\langle U^+(t)U(\tau)\rangle_{l,L} + \mathrm{c.c.}\right)\right] \\
\times \frac{\partial^2}{\partial\alpha\partial\alpha^*}\right\}P(\alpha, \alpha^*, t).$$

Equation (12), which is linear in the field amplitude, describes the laser action the below the threshold and allows us to determine the lasing threshold by equating the loss rate to the gain G:

$$\kappa = G, \\
G \equiv \frac{1}{2}g^2 N\sigma_0 \quad (13) \\
\times \left(\int_0^\infty d\tau\gamma_\perp \exp[(i\Delta-\gamma_\perp)\tau]\langle U^+(t)U(\tau)\rangle_{l,L} + \mathrm{c.c.}\right).$$

The coefficient at the second-order derivative with respect to the amplitude variable in (12)

$$2\kappa\bar{N} + \frac{g^2 N(1+\sigma_0)}{2\gamma_\perp}\left(\int_0^\infty d\tau\gamma_\perp \\
\times \exp[(i\Delta-\gamma_\perp)\tau]\langle U^+(t)U(\tau)\rangle_{l,L} + \mathrm{c.c.}\right) \equiv D \quad (14)$$

is the diffusion constant.

Recall that $\bar{N} = [\exp(\omega/k_BT) - 1]^{-1}$ in (14) is the occupation number of the photon mode at temperature T and N is the number of impurity particles.

## 3. THRESHOLD CONDITIONS OF LASING AT THE VIBRONIC REPLICAS OF THE PURELY ELECTRONIC LINE

By using expression (11) for the correlation function $\langle U^+(t)U(\tau)\rangle_{l,L}$, the gain can be represented in the explicit form

$$\begin{aligned} G &= \frac{g^2 N \sigma_0}{\gamma_\perp} \exp[-\xi^2(2\bar{n}+1)] \\ &\times \sum_{p=-\infty}^{\infty} \sum_{r,s=0}^{\infty} \prod_k \xi^{2(r+s)} (\bar{n}+1)^r \bar{n}^s \gamma_\perp \\ &\times [\gamma_\perp + (r+s)\alpha + \xi^2(2\bar{n}+1)\alpha] \\ &\times \{r! s! [\varepsilon - \omega - (r-s)\nu - p\omega_k]^2 \\ &+ [\gamma_\perp + (r+s)\alpha + \xi^2(2\bar{n}+1)\alpha]^2 \}^{-1} \\ &\times \exp[-w(T)] I_p\left(2\frac{\lambda_k^2}{\omega_k^2}\sqrt{\bar{n}_k(\bar{n}_k+1)}\right)\left(\frac{\bar{n}_k+1}{\bar{n}_k}\right)^{p/2}, \end{aligned} \qquad (15)$$

where $\exp[-w(T)]$ is the Debye–Waller factor, which is related to the total Stokes loss

$$\sum_k \left(\frac{\lambda_k}{\omega_k}\right)^2 (2\bar{n}_k + 1)$$

caused by phonons, and $\exp[-\xi^2(2\bar{n}+1)]$ is the Debye–Waller factor related to local vibrations; $I_p$ is the Bessel function of the imaginary argument of the first kind; $\varepsilon$ is the purely electronic transition energy; $\omega$ and $\kappa$ are the frequency and the loss rate of a photon mode in the cavity, respectively; $\nu$ and $\alpha$ are the frequency and the decay of a local (intramolecular) vibration; $\omega_k$ is the frequency of the $k$th phonon; and

$$\gamma_\perp = \gamma_\parallel/2 + \Gamma(T) \qquad (16)$$

is the relaxation rate of the induced polarization, where $\gamma_\parallel = \gamma_{12} + \gamma_{21}$.

Consider first of all several particular cases concerning the threshold condition (13), (15). First, in the absence of the adiabatic interaction of the ground and excited electronic states of the optical centers with local vibrations and phonons ($\xi \approx 0$, $\lambda_k \approx 0$), the threshold condition (13), (15) coincides with the usual one [13],

$$\kappa = \frac{g^2 N \sigma_0}{\gamma_\perp}. \qquad (17)$$

Second, in the case of a very strong coupling of optical centers with the vibrational modes ($\xi^2 \gg 1$, $\lambda_k^2/\omega_k^2 \gg 1$), as well as at high temperatures, the gain $G$ becomes so small that the equality (13) cannot be satisfied, and lasing is impossible. Third, for the purely electronic transition ($r = s = p = 0$), in the case of the exact resonance ($\varepsilon - \omega = 0$), we have the threshold condition [14]

$$\kappa = \frac{g^2 N \sigma_0}{\tilde{\gamma}_\perp} \exp[-\tilde{w}(T)], \qquad (18)$$

where $\tilde{w}(T) = w(T) + \xi^2(2\bar{n}+1)$ and $\tilde{\gamma}_\perp = \gamma_\perp = \gamma_\perp + \xi^2(2\bar{n}+1)\alpha$. The threshold condition (18) corresponds to the case when lasing occurs within the zero-phonon line, i.e., photons interact with the impurity crystal in such a way that the vibrational states of both local modes and phonons do not change.

More typical is the situation when the interaction of light with an impurity crystal does not change the vibrational state of the crystal lattice (p = 0), but the resonance condition $\omega = \varepsilon - (r - s)\nu$ is simultaneously satisfied, so that one of the vibrational replicas of the purely electronic line is observed, which corresponds to the vibronic transition from the $s$th vibrational level in the excited electronic state (when $s$ vibrational quanta are annihilated) to the $r$th vibrational level of the ground electronic state (when $r$ vibrational quanta are created). The lasing threshold for such an individual vibronic transition can be written in a rather simple form

$$\begin{aligned} \kappa &= \frac{g^2 N \sigma_0}{\gamma_\perp} \exp[-\xi^2(2\bar{n}+1)] \\ &\times \frac{\gamma_\perp \xi^{2(r+s)} (\bar{n}+1)^r \bar{n}^s}{r! s! [\gamma_\perp + \xi^2(2\bar{n}+1)\alpha + (r+s)\alpha]} \\ &\times \exp[-w(T)] \prod_k I_0\left(2\frac{\lambda_k^2}{\omega_k^2}\sqrt{\bar{n}_k(\bar{n}_k+1)}\right). \end{aligned} \qquad (19)$$

which is convenient for estimates.

Expression (19) can be used for determining the lasing thresholds for any of the vibronic transitions, neglecting their interrelation [in the general case, expressions (13) and (15) should be used]. If sufficiently high gains are simultaneously achieved for a whole series of vibronic transitions, so that the corresponding threshold conditions are realized, then, due to a distribution of the field modes in the cavity (because the emission line has a finite width), the resonance can be achieved at several transitions rather than at one transition, and tunable lasing became possible.

The product of the zero-order Bessel functions in expression (19) changes with temperature (or $\bar{n}_k$) slower than the Debye–Waller factor $\exp[-w(T)]$, and for $(\lambda_k/\omega_k)^2 \sqrt{\bar{n}_k(\bar{n}_k+1)} \ll 1$, this product can be set equal to unity because $I_0(x) \approx 1$. Because the intramolecular frequency is much higher than the phonon frequency, the temperature dependence of the gain for different vibronic transitions [see the right-hand side of equation (19)] will be mainly determined by the Debye–Waller factor $\exp[-w(T)]$. It also follows from expression (19) that as the strength of interaction (i.e., the Stokes loss $\xi^2$) of the optical centers with an intramolecular vibrations increases, the maximum gain is achieved for the longer-wavelength vibronic transitions.

## 4. LASING IN THE REGION OF THE STRUCTURELESS PHONON SIDEBAND

If the optical spectrum of a crystal consists of only a zero-phonon line and a structureless phonon sideband, i.e., the coupling of electronic states with high-frequency local vibrations is negligible (or the latter are absent at all), then we should set $\xi = \nu = \alpha = 0$ in the above expressions and, in particular, in expression (15) for the gain. For the estimates that will be performed below, it is convenient to replace the phonon spectrum of the crystal by an effective frequency $\omega_0$ (the Einstein approximation). Moreover, if lasing takes place within a narrow region of the phonon sideband, the condition of the exact resonance

$$\varepsilon - \omega - p\omega_0 = 0 \qquad (20)$$

can be reasonably imposed. Then, the corresponding gain, taking into account expression (15), can be rewritten in the form

$$G(\omega = \varepsilon - p\omega_0; \lambda_0; T) = \frac{g^2 N \sigma_0}{\gamma_\perp(T)} f(\lambda_0; T), \quad (21)$$

$$f(\lambda_0; T) = \exp\left\{-\frac{\lambda_0^2}{\omega_0^2}[2\bar{n}(\omega_0) + 1]\right\}$$
$$\times I_p\left(2\frac{\lambda_0^2}{\omega_0^2}\sqrt{\bar{n}(\omega_0)(\bar{n}(\omega_0)+1)}\right)\left(\frac{\bar{n}(\omega_0)+1}{\bar{n}(\omega_0)}\right)^{p/2}. \quad (22)$$

The arguments of the function $G$ (21) show that the gain depends on the electron–phonon coupling constant $\lambda_0$ and the temperature $T$ of the crystal lattice, as well as that an elementary event of emission of a quantum of the electromagnetic field of frequency $\omega$ during lasing is accompanied by the creation of $p$ phonons with frequency $\omega_0$ due to the resonance condition (20).

The linear Fokker–Planck equation (12) is sufficient for the determination of the lasing threshold and the diffusion constant, whereas the real amplitude of the laser-field oscillations can be found only with the help of the nonlinear theory, i.e., taking into account in the derivation of the master equation for the density matrix the terms at least up to the fourth order inclusive in the interaction $L_{af}$ (or $H_{af}$). However, if the resonance condition (20) is satisfied, the field amplitude can be found by modifying equations (14)–(16) from paper [14] by making the substitution $g\exp[-S(T)/2] \longrightarrow gf^{1/2}(\lambda_0; T)$. As a result, the polarization $P$, the inversion $\Delta$, the amplitude $a$, and the average number of phonons $n = \langle a^+ a \rangle = a^2$ will be determined, in the average field approximation, by the set of equations

$$\dot{P} = 2ga\Delta f^{1/2} - \gamma_\perp P,$$
$$\dot{\Delta} = -2gaPf^{1/2} - \gamma_\parallel(\Delta - \sigma_0 N/2),$$
$$\dot{a} = 2g\Delta f^{1/2} - \kappa a, \qquad (23)$$
$$\dot{n} = 2gaPf^{1/2}.$$

Assuming $\dot{P} = \dot{\Delta} = 0$, we find the evolution equation for the field amplitude generated in the region of the phonon sideband at the frequency $\omega = \varepsilon - p\omega_0$:

$$\dot{a} = (G - \kappa)a - \frac{4\kappa}{N\gamma_\parallel \sigma_0} Ga^3. \qquad (24)$$

For $G > \kappa$, we can find from (24) the average number of photons in the stationary regime ($\dot{a} = 0$):

$$n = a^2 = \frac{N\gamma_\parallel}{4\kappa}(\sigma_0 - \sigma_0 \kappa G^{-1}). \qquad (25)$$

In this case, the laser radiation intensity is

$$I = \omega \dot{n} = 2g\omega aPf^{1/2} = 2\kappa n, \qquad (26)$$

or, after the substitution of (25) into (26),

$$I = \frac{N\gamma_\parallel}{2}\left[\sigma_0 - \frac{\kappa\gamma_\perp(T)}{g^2 N}f^{-1}(\lambda_0; T)\right], \qquad (27)$$

where the relation (21) between $G$ and $f$ is taken into account.

## 5. APPLICATION OF THE THEORY TO THE DESCRIPTION OF LASING IN A $Mg_2SiO_4$ : $Cr^{4+}$ SINGLE CRYSTAL

The theory of a vibronic laser developed above can be compared with experiments on lasing observed in the phonon sideband in the optical spectrum of a $Mg_2SiO_4$ : $Cr^{4+}$ forsterite single crystal. This laser produces stable quasi-continuous emission in the near IR region with an average output power of above 1 W [3, 4]. The pump frequency, the purely electronic transition frequency, and the lasing frequency were 9276, 9158, and 8097 cm$^{-1}$, respectively. Because no decrease in the quantum yield of luminescence was observed in the temperature range between 300 and 400 K of interest for applications, i.e., the nonradiative relaxation was absent, this means that the vibrational potential curves of the atoms in the configuration space in the ground and metastable electronic states should not intersect (which was noted in [4]).

In [4], the temperature dependence of the energy efficiency $\eta$ (the ratio $I/I_p$ between the output and pump powers) of a $Mg_2SiO_4$ : $Cr^{4+}$ laser was measured. Expression (27) can be transformed to describe the experimental situation. Indeed, let the excess over the lasing

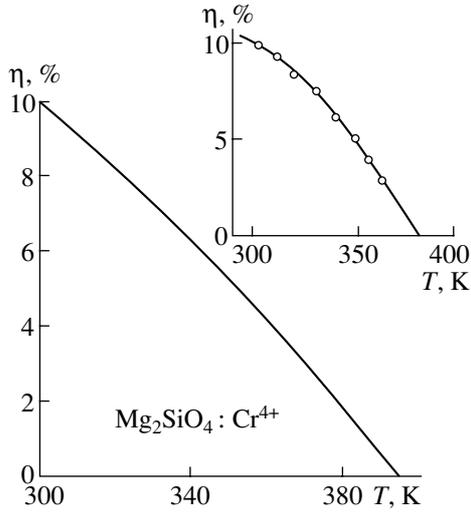

Temperature dependence of the energy efficiency of lasing η in the region of the phonon sideband. The insert shows the experimental temperature dependence of the efficiency of a chromium-doped forsterite laser [4].

threshold at some initial temperature $T_0$ is $\Delta$ and η equals $\eta_0$ (in per cent), then, using (27), we can write

$$\eta(T) = \frac{I(T)}{I_p} = \eta_0 \Delta$$
$$\times \left[ 1 - \frac{1}{\Delta} \frac{\bar{n}(\omega_0, T)(\bar{n}(\omega_0, T) + 1)}{\bar{n}(\omega_0, T)(\bar{n}(\omega_0, T) + 1)} \frac{f(\lambda_0, \omega_0; T_0)}{f(\lambda_0, \omega_0; T)} \right]. \quad (28)$$

Expression (28) takes into account that, at sufficiently high temperatures, for example, room temperature, the rate $\gamma_\parallel$ of energy relaxation is much lower than the dephasing contribution $\Gamma(T)$ into (16), so that in fact we have

$$\gamma_\perp(T) \approx \chi^2 \bar{n}(\omega_0, T)[\bar{n}(\omega_0, T) + 1] \quad (29)$$

[see expressions (4) and (16)]. The function $f(\lambda_0, \omega_0; T)$ is defined in (22). Thus, given the reference points $T_0$, $\Delta$, and $\eta_0$, we can express the energy efficiency of laser (28) only in terms of the crystal-lattice parameters, such as the effective frequency $\omega_0$, temperature $T$, and the electron–phonon coupling strength $\lambda_0$. However, note that the effective frequency $\omega_0$ in expression (29) for $\gamma_\perp$ or in expression (28) for $\bar{n}$ can be in the general case different from this quantity appearing as an argument of the function $f$ in (28).

The excess over the lasing threshold at $T_0 = 300$ K in the experiment [4] was $\Delta = 2.2$ and the value of $\eta_0$ was 10%. The temperature was measured with an external heater. The experimental results are presented in the insert in the figure. One can see that the energy efficiency of the laser decreases by 25% at T < 330 K and lasing disappears at 380 K.

In the calculation of η(T) from expression (28), we will assume that $\omega_0$ is equal to the difference between the purely electronic transition frequency 9158 cm$^{-1}$ and the lasing frequency 8097 cm$^{-1}$, i.e., to 1061 cm$^{-1}$, and, because in this case only one-quantum transitions occur, we will set $p = 1$ in expression (22) for $f$ and $\lambda_0^2/\omega_0^2 = 1$. The result of calculations is presented in the figure. The theory adequately describes the temperature dependence η(T), both qualitatively and quantitatively, despite the fact that the phonon spectrum of the crystal was represented by a very simplified one-oscillator model.

## 6. CONCLUSIONS

In this paper, some aspects of the operation of a vibronic laser below the threshold have been studied. It has been shown that the threshold conditions of lasing at vibronic transitions depend on the strength of the interaction of optical centers with a local (intramolecular) vibration that forms the vibronic spectrum—a series of vibrational replicas of the purely electronic line. These threshold conditions also depend on the crystal lattice temperature. As the Stokes loss per local vibration increases, the maximum value of the gain is achieved at the longer-wavelength vibronic transitions, while the increase in the crystal temperature reduces the gain at all the vibronic transitions virtually in the same degree. The theory can be readily generalized to the spectrum containing a structureless phonon sideband and well agrees with the experimental temperature dependence of the output of power of a $Mg_2SiO_4 : Cr^{4+}$ forsterite laser.